\documentclass[aps,prb,twocolumn, amsmath, amssymb, superscriptaddress]{revtex4-2}
\usepackage{graphicx}
\usepackage{bm}
\usepackage{amsmath}
\usepackage{float}

\begin{document}

\title{Anisotropy of Yu-Shiba-Rusinov states in NbSe$_2$}

\author{Mateo Uldemolins}
\email[]{mateo.uldemolins@dipc.org}
\affiliation{Donostia International Physics Center (DIPC), 20018, Donostia–San Sebasti\'an, Spain}
\affiliation{Universit\'e Paris-Saclay, CNRS, Laboratoire de Physique des Solides, 91405, Orsay, France}
\author{Freek Massee}
\affiliation{Universit\'e Paris-Saclay, CNRS, Laboratoire de Physique des Solides, 91405, Orsay, France}
\author{Tristan Cren}
\affiliation{Sorbonne Universit\'e, CNRS, Institut des Nanosciences de Paris, UMR7588, F-75252 Paris, France}
\author{Andrej Mesaros}
\affiliation{Universit\'e Paris-Saclay, CNRS, Laboratoire de Physique des Solides, 91405, Orsay, France}
\author{Pascal Simon}
\affiliation{Universit\'e Paris-Saclay, CNRS, Laboratoire de Physique des Solides, 91405, Orsay, France}

\date{\today}

\begin{abstract}
The spatial structure of in-gap Yu-Shiba-Rusinov (YSR) bound states induced by a magnetic impurity in a superconductor is the essential ingredient for the possibility of engineering collective impurity states. Recently, a saddle-point approximation [Phys.~Rev.~B 105, 144503] revealed how the spatial form of a YSR state is controlled by an anisotropic exponential decay length, and an anisotropic prefactor, which depends on the Fermi velocity and Fermi-surface curvature. Here we analyze scanning tunnel microscope (STM) data on YSR states in NbSe$_2$, focusing on the key issue that the exponential decay length predicted theoretically from the small superconducting gap is much larger than the observed extent of YSR states. We confirm that the exponential decay can be neglected in the analysis of the anisotropy. Instead, we extract the anisotropic prefactor directly from the data, matching it to the theoretical prediction, and we establish that the theoretical expression for the prefactor alone captures the characteristic flower-like shape of the YSR state. Surprisingly, we find that up to linear order in the superconducting gap the anisotropic prefactor that determines the shape of YSR states is the same as the anisotropic response to the impurity in the underlying normal metal. Our work points out the correct way to analyze STM data on impurities in small-gap superconductors, and reveals the importance of the normal band structure's curvature and Fermi velocity in designing multi-impurity in-gap states in superconductors.
\end{abstract}

\maketitle

\section{Introduction}

The local response to single magnetic impurities in conventional superconductors appears most directly in the form of pairs of in-gap excitations localized on the impurity known as Yu-Shiba-Rusinov (YSR) states \cite{rusinovSuperconductivityParamagneticImpurity1969,shibaClassicalSpinsSuperconductors1968,yuluhBOUNDSTATESUPERCONDUCTORS1965}. YSR states are nowadays routinely probed by scanning tunneling microscopy (STM) experiments \cite{heinrichSingleMagneticAdsorbates2018} and, while an old problem, they have attracted a renewed wave of interest owing to their potential to extract information on the properties of the host \cite{kaladzhyan2016, perrin2020, Sukhachov_2023} and study magnetism \cite{choiMappingOrbitalStructure2017, choiInfluenceMagneticOrdering2018} and spin-transport phenomena \cite{huangTunnellingDynamicsSuperconducting2020, huangSpindependentTunnelingIndividual2021,schneiderAtomicscaleSpinpolarizationMaps2021, Vaxevani_2022} at the atomic scale. At the same time, much experimental and theoretical effort has been devoted to designing complex structures in real space (e.g., impurity chains or lattices) \cite{nadj-perge2013, pientka2013, braunecker2013, nakosaki2013, Nadj_Perge_2014, rontynen2015, li2016two, Jeon_2017, ruby2017exploring, kim2018toward, palacio2019atomic, Liebhaber_2022, farinacci2023yushibarusinov, Soldini_2023}, which are suggested to realize topological states of matter harboring Majorana zero modes \cite{yazdaniHuntingMajoranas2023}. Understanding the spatial structure of YSR states is key for developing this program.

In an isotropic $d$-dimensional system, the local density of states (LDOS) of YSR states far away from the impurity behaves according to $\sim e^{-r/\xi}/r^{d-1}$, with the superconducting coherence length controlling the exponential decay, $\xi \sim \frac{v_{\mathbf{F}} \hbar}{\Delta} \sim \xi_{\mathrm{SC}}$. The dependence of the power-law on the dimensionality of the substrate has been evidenced by STM measurements on two-dimensional superconductors ($d=2$), e.g., NbSe$_2$, where the in-gap LDOS extends over several nanometers \cite{menardCoherentLongrangeMagnetic2015, kimLongrangeFocusingMagnetic2020, thupakulaCoherentIncoherentTunneling2022}, thereby rendering this class of materials a promising platform for constructing collective impurity states \cite{Kezilebieke_2018, liebhaberYuShibaRusinov2020, ptok2017, sticlet2019, rutten2024wave}. In general, the measured in-gap LDOS oftentimes displays a marked anisotropy on length scales largely surpassing the typical impurity size and hence attributed to the lower symmetry of the band structure \cite{kimLongrangeFocusingMagnetic2020, ortuzarYuShibaRusinovStates2D2022}. The theory of spatial structure of YSR states is hence intricate, requiring a blend of information about the substrate's band structure, pairing function, and impurity coupling.

Recently, the theory of the quasiparticle focusing effect (QFE) \cite{weismannSeeingFermiSurface2009, lounisTheoryRealSpace2011} was extended to superconducting substrates by some of the authors \cite{uldemolinsQuasiparticleFocusingBound2022}, leading to the following insights in the limit of weak pairing, $\Delta/\varepsilon_{\mathrm{F}}\ll1$, and at long distances ($r k_{\mathrm{F}} \gg 1$): 
\begin{enumerate}
    \item The QFE persists, i.e., the response to the impurity in a given real-space direction is controlled by momenta (denoted \textit{critical}) on the Fermi surface at which the Fermi velocity lies parallel or antiparallel to that real-space direction. This selectivity in reciprocal space controls the spatial anisotropy of the YSR states; in particular, YSR states are focused in directions perpendicular to flatter sections of the Fermi surface.
    \item Specifically to gapped superconductors, the in-gap YSR LDOS has an additional exponentially decaying factor with angularly dependent exponential decay length, apart from a prefactor and oscillations, all related to the geometry of the Fermi surface through a simple analytical expression. Importantly, the prefactor and exponential decay length are essentially always in phase, i.e.~they agree on the direction along which the LDOS is focused.
\end{enumerate}
This theory describes very accurately the anisotropy of the characteristic exponential decaying length of YSR states simulated on a lattice \cite{uldemolinsQuasiparticleFocusingBound2022} and successfully predicts the main anisotropy directions of YSR states observed in NbSe$_2$. However, the tight-binding description of the Nb-derived bands at the Fermi level \cite{rahnGapsKinksElectronic2012} in combination with the small superconducting gap ($\Delta \sim 1$ meV \cite{rodrigo_2004, kuzmanovicTunnelingSpectroscopyFewmonolayer2022}), yields an in-gap LDOS with large exponential decay length ($\xi_{\mathrm{TB}} \sim 100$ nm) \cite{menardCoherentLongrangeMagnetic2015, liebhaberYuShibaRusinov2020, sticlet2019, rutten2024wave} that widely surpasses the extent of the observed YSR states, as well as the superconducting coherence length reported in this material ($\xi^{\mathrm{exp}}_{\mathrm{SC}} \sim 10$ nm \cite{sanchez_1995, nader_2014, mourachkine2004determination}). Therefore, there is a problem in the natural interpretation that the anisotropy of the coherence length in NbSe$_2$ fundamentally controls the shape of the YSR LDOS, and theoretical radial fitting of an exponential decay is questionable.

In this work, we show that the angular dependence of the LDOS prefactor accounts for the anisotropy of YSR states in this material, while the radial fitting of the LDOS can be circumvented, i.e., the (anisotropic) exponential decay can be neglected. This evokes the QFE in normal metals \cite{weismannSeeingFermiSurface2009, lounisTheoryRealSpace2011} where standard Friedel oscillations decay algebraically and lack the exponential factor. In fact, our analytical expression of the prefactor has the same functional form, in terms of the Fermi velocity and curvature of the Fermi surface, as the LDOS in the case of a normal metal. A difference in principle arises when this expression is evaluated at the critical momenta, which vary as the band structure of the metal changes into the one of the superconductor. Surprisingly, we find that up to linear order in $\Delta/\varepsilon_{\mathrm{F}}\ll1$ the prefactor of the impurity-induced LDOS in the superconductor exactly equals the one in the parent normal metal. We show that this prefactor can be reliably extracted from STM data on NbSe$_2$ and that the characteristic flower-shaped YSR LDOS is faithfully modeled by the prefactor alone. We also find that introducing the effect of the charge density wave (CDW) on the bandstructure does not play a significant role in the YSR LDOS.

The rest of the paper is organized as follows: we start by presenting the STM data in Sec.~\ref{sec:res}. Then, in Sec.~\ref{ssec:real_space}, we briefly recall the main analytical expressions derived in \cite{uldemolinsQuasiparticleFocusingBound2022} and we  analyze the spatial anisotropy of the LDOS prefactor. This analysis is complemented in Sec.~\ref{ssec:fourier_space}, where we study the Fourier transform of the YSR states. We conclude with a short discussion in Sec.~\ref{sec:conclusions}. Additional details concerning the experiment and the theory expressions are presented in the Appendixes.

\section{Results}
\label{sec:res}

\begin{figure}
 \includegraphics[width=\columnwidth]{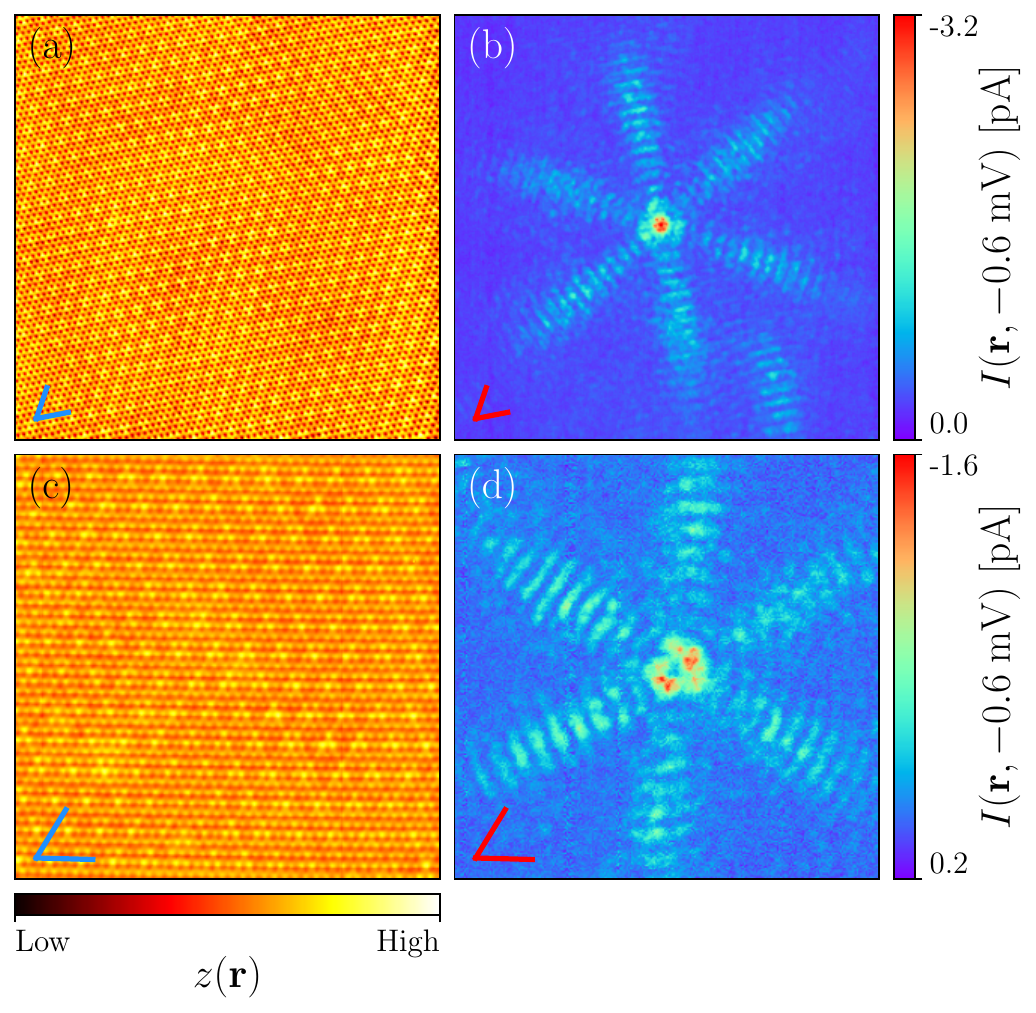}
 \caption{Spatial data sets 1 and 2. (a) Atomically resolved constant-current image of the Se surface of 2H-NbSe$_2$, $V_{\mathrm{set}}=4$ mV and $I_{\mathrm{set}}=160$ pA. (b) In-gap current recorded at $V=-0.6$ mV showing a sixfold YSR state. (c),(d) The same as (a),(b) for another YSR state, with $V_{\mathrm{set}}=4.2$ mV, $I_{\mathrm{set}}=100$ pA, and $V = -0.6$ mV, respectively. The lines at the bottom-left corner indicate the crystallographic axes of 2H–NbSe$_2$. Their length corresponds to 2 nm in both data sets.}
 \label{fig:spatial_data}
\end{figure}

We consider STM data on 2H-NbSe$_2$ crystals containing a few tens of ppm of magnetic atoms measured at 300 mK with a metallic tip (for additional details concerning the fabrication and measurement conditions, see Appendix~\ref{appExp}). YSR bound states manifest as an in-gap peak in the differential-conductance spectrum at the impurity location (see Appendix~\ref{appExp}). In Fig.~\ref{fig:spatial_data}, we present atomically resolved data around two independent impurities on the same sample. The constant-current image [Figs.~\ref{fig:spatial_data}(a) and \ref{fig:spatial_data}(c)] shows the triangular geometry of the Se lattice without any visible imperfection, implying that the defect giving rise to the YSR state is most likely an atom substitution in the Nb layer. The corresponding current images recorded before the onset of the superconducting coherence peak [Figs.~\ref{fig:spatial_data}(b) and \ref{fig:spatial_data}(d)] show a marked six-fold anisotropy that constitutes an excellent example of the QFE on a superconductor.

\subsection{Characterization of the LDOS prefactor in real space}
\label{ssec:real_space}
Let us begin by briefly recalling the theoretical framework to describe the QFE on superconductors. We model the impurity-substrate system as a classical spin on an $s$-wave superconductor. 

The Hamiltonian reads,
\begin{subequations}
\begin{align}
 &H = H_0+H_{\mathrm{imp}},\label{eq:h_tot} \\
 &H_0 = \sum_{\bm{k}\sigma} \varepsilon_{\bm{k}\sigma} c^\dagger_{\bm{k} \sigma} c_{\bm{k} \sigma} + \Delta \sum_{\bm{k}}  \; c^\dagger_{\bm{k} \uparrow} c^\dagger_{\bm{-k} \downarrow} + \mathrm{h.c.},\label{eq:h0} \\
 &H_{\mathrm{imp}} = -J \left( c^\dagger_{\bm{r}_0\uparrow} c_{\bm{r}_0\uparrow} - c^\dagger_{\bm{r}_0\downarrow} c_{\bm{r}_0\downarrow} \right),\label{eq:h_imp}
\end{align}
\end{subequations}
where $\varepsilon_{\bm{k}}$ is the spin-degenerate normal energy dispersion of the substrate, $\Delta$ is the BCS superconducting gap and $J$ is the amplitude of the magnetic exchange coupling between the superconducting electrons and the spin of the impurity at $\bm{r}_0$. Specifically, $\varepsilon_{\bm{k}}$ is an effective fifth-nearest neighbors tight-binding energy dispersion on a triangular lattice that describes one of the two lowest-lying Nb 4d bands of 2H-NbSe$_2$ \cite{rahnGapsKinksElectronic2012} (see Appendix~\ref{appTB}). This description neglects interlayer coupling of the bulk crystal and assumes that the magnetic impurity couples to one band only \cite{menardCoherentLongrangeMagnetic2015, liebhaberYuShibaRusinov2020}. It also neglects spin-orbit coupling \cite{xiIsingPairingSuperconducting2016}; however, it provides a faithful description of the geometry of the Fermi surface and band structure of the substrate, which is the crucial element to characterize the QFE. 

An important fact about the band structure of niobium diselenide is that it develops CDW at $T_{\mathrm{CDW}} = 33K$, which coexists with superconductivity ($T_{\mathrm{c}}=7.2K$) at low temperatures \cite{ugeda_cdw_2015, dreher_2021, flicker_2015}. Although density functional theory indicated a quite large deformation of the Fermi surface due to the CDW \cite{Silva-Guillen_2016}, for the particular case of 2H-NbSe$_2$ the ARPES data show that the shape and size of the Fermi contours are largely preserved well below the CDW transition temperature \cite{rahnGapsKinksElectronic2012}. Therefore, neglecting the effect of CDW on the long-range spatial properties of the YSR LDOS should be a reasonable simplification as well. For completeness, in Appendix~\ref{app:CDW} we discuss in detail the effect of the CDW within our approach. We hence confirm that although the CDW introduces interesting effects in principle, for the effective parameter values derived from ARPES there is no change to any of our main conclusions. Hence for simplicity we neglect the presence of CDW in the rest of the main text.

In addition, Eq.~\eqref{eq:h_imp} assumes a fully local and isotropic scatterer, thereby neglecting the magnetic ion's $d$-orbital structure and the adsorption site's symmetry. These add to the YSR-state wave function some degree of anisotropy \cite{choiMappingOrbitalStructure2017,rubyOrbitalPictureYuShibaRusinov2016}; however, the observed six-fold anisotropy of the LDOS [Figs.~\ref{fig:spatial_data}(b) and \ref{fig:spatial_data}(d)] occurs on length scales of several nanometers that largely surpass the typical orbital size; therefore, it can be safely ascribed to the QFE and justifies the model's simplification (see also Appendix~\ref{app:CDW}).

The LDOS at a distance $\bm{r}$ from the impurity and at the bound-state energy $E_{\mathrm{S}}$ is computed by means of a saddle-point approximation on the bare propagator valid in the far-field limit ($rk_{F, \mathrm{min}} \gg 1$, with $k_{F, \mathrm{min}}$ the minimum Fermi wave vector of the Fermi contour). By further taking the small-gap limit ($\Delta / \varepsilon_{\mathrm{F}} \ll 1$), the LDOS takes the approximate form
\begin{equation}
 \label{eq:rho}
 \rho(\bm{r}) \sim \frac{1}{r}\sum_{j,j'}\Gamma_{j,j'}(\theta_{\bm{r}}) e^{-r/\xi_{j,j'}(\theta_{\bm{r}})}f[\bm{k}^{\pm}_{j,j'}(\theta_{\bm{r}})\cdot\bm{r}],
\end{equation}
where the angular-dependent prefactor reads
\begin{equation}\label{eq:Gamma}
\Gamma_{j,j'}(\theta_{\bm{r}}) = \frac{1}{|\nabla\varepsilon_{\widetilde{\bm{k}}_j(\theta_{\bm{r}})}|\sqrt{\kappa_{\widetilde{\bm{k}}_j(\theta_{\bm{r}})}}|\nabla\varepsilon_{\widetilde{\bm{k}}_{j'}(\theta_{\bm{r}})}|\sqrt{\kappa_{\widetilde{\bm{k}}_{j'}(\theta_{\bm{r}})}}},
\end{equation}
with $|\nabla\varepsilon_{\widetilde{\bm{k}}_j(\theta_{\bm{r}})}|$ the Fermi velocity and $\kappa_{\widetilde{\bm{k}}_j(\theta_{\bm{r}})}$ the curvature of the Fermi contour evaluated at $\widetilde{\bm{k}}_j(\theta_{\bm{r}})$. The critical momenta $\widetilde{\bm{k}}_j(\theta_{\bm{r}})$ are solutions to the saddle-point equations in the small-gap limit, namely, points on the Fermi contour for which the Fermi velocity is parallel to $\bm{r}$,
\begin{equation}
 \label{eq:saddle_eq}
  \frac{\nabla\varepsilon_{\widetilde{\bm{k}}_j(\theta_{\bm{r}})}}{|\nabla\varepsilon_{\widetilde{\bm{k}}_j(\theta_{\bm{r}})}|} = \bm{\hat{r}},
 \end{equation}
hence their dependence on the polar angle $\theta_{\bm{r}}$ which controls the spatial anisotropy of the LDOS. We recall that the gapped nature of the superconductor yields complex solutions to the saddle-point equations \cite{uldemolinsQuasiparticleFocusingBound2022}. However, up to and including the linear order in $\Delta/\varepsilon_{\mathrm{F}}$, these solutions [i.e. the vectors $\widetilde{\bm{k}}_j(\theta_{\bm{r}})$] are real and positioned on the Fermi contour of the underlying metal. Therefore, the prefactor $\Gamma_{j,j'}(\theta_{\bm{r}})$ in Eq.~\eqref{eq:Gamma} in the small-gap limit is simply determined by the band structure of the normal metal (see Appendix~\ref{appLDOS}). The exponential decay length $\xi_{j,j'}(\theta_{\bm{r}})$ depends on the Fermi velocity at $\widetilde{\bm{k}}_j(\theta_{\bm{r}})$ and $\widetilde{\bm{k}}_{j'}(\theta_{\bm{r}})$,
and $f[\bm{k}^{\pm}_{j,j'}(\theta_{\bm{r}})\cdot\bm{r}]$ is a linear combination of oscillating functions with typical characteristic inverse lengths $\bm{k}^{\pm}_{j,j'}(\theta_{\bm{r}}) = \widetilde{\bm{k}}_j(\theta_{\bm{r}}) \pm \widetilde{\bm{k}}_{j'}(\theta_{\bm{r}})$. The indices $j,j'=1,\dots,N$ label the various solutions to Eq.~\eqref{eq:saddle_eq}. The form in Eq.~\eqref{eq:rho}, derived in Appendix~\ref{appLDOS}, is more compact than the form first derived in Ref.~\cite{uldemolinsQuasiparticleFocusingBound2022}, as it pairs up the two contributions from the wavevectors whose respective Fermi velocities are parallel and antiparallel to the observation direction [the latter satisfying Eq.~\eqref{eq:saddle_eq} with $\hat{\bm{r}}\rightarrow-\hat{\bm{r}}$]. The Fermi contour of NbSe$_2$ consists of three disconnected pockets centered at $\Gamma$, $K$, and $K'$ and, therefore, $N=3$. 

To analyze the spatial anisotropy of the STM data we consider radial averaging in annuli around the impurity and hence introduce the following quantity:
\begin{equation}
    \label{eq:int_rho_exp}
    I_{\mathrm{int}}(\theta_{\bm{r}}) = \frac{1}{M}\sum_{\bm{R}_m\in M(\theta_{\bm{r}})}|\bm{R}_m|\cdot I(\bm{R}_m),
\end{equation}
where $I(\bm{R}_m)$ is the measured current at point $\bm{R}_m$ sitting at a distance $|\bm{R}_m|$ from the impurity, while $M(\theta_{\bm{r}})$ is the set of $M$ points belonging to an angular sector centered on the impurity, having infinitesimal angular width $d\theta$, an inner radius $R_0$, and outer radius $R_0+\delta$. As long as $R_0$ is sufficiently large to make the asymptotic approximation reasonable ($R_0 \gg 1/k_{\mathrm{F}}^{\mathrm{min}}\sim 0.2 \; \mathrm{nm}$ for the Fermi contour under consideration), we can relate $I_{\mathrm{int}}(\theta_{\bm{r}})$ to the radially integrated LDOS,
\begin{equation}
    \label{eq:int_rho_theo}
     I_{\mathrm{int}}(\theta_{\bm{r}})\sim\rho_{\mathrm{int}}^{R_0,\delta}(\theta_{\bm{r}}) = \frac{1}{\delta}\int_{R_0}^{R_0+\delta}\textrm{d}r r \rho_{\mathrm{max}}(\bm{r}),
\end{equation}
where $\rho_{\mathrm{max}}(\bm{r})$ is the term in Eq.~\eqref{eq:rho} with dominant prefactor, $\Gamma_{\mathrm{max}}(\theta_{\bm{r}}) = \mathrm{max}\{\Gamma_{j,j'}(\theta_{\bm{r}})\}$. This simplification is legitimate as the various prefactors $\Gamma_{j,j'}(\theta_{\bm{r}})$ are essentially in phase (see Appendix~\ref{appTB} for further details). We can further simplify Eq.~\eqref{eq:int_rho_theo} by making the following assumptions:
\begin{itemize}
    \item The integration range $\delta$ is larger than the typical oscillation length; therefore, $f[\bm{k}^{\pm}_{j,j'}(\theta_{\bm{r}})\cdot\bm{r}]$ is averaged out and can be safely ignored.
    \item The exponential decay length is sufficiently large so that $e^{-r/\xi(\theta)}\sim 1$ over the integration range. This demands $R_0+\delta \leq \xi^{\mathrm{exp}}_{\mathrm{SC}}\sim 10 \; \mathrm{nm} \ll \mathrm{min}\{\xi_{j,j'}(\theta_{\bm{r}})\}$. This approximation also relies on the fact that the angular dependence of the prefactor and the exponential decay length are in phase and therefore there are not any competition effects.
\end{itemize}
Under these assumptions $\rho_{\mathrm{int}}^{R_0,\delta}(\theta_{\bm{r}})$ becomes independent of $\delta$ and, further, we have
\begin{equation}
    \rho_{\mathrm{int}}^{R_0,\delta}(\theta_{\bm{r}}) \sim \Gamma_{\mathrm{max}}(\theta_{\bm{r}}).
\end{equation}

\begin{figure}[H]
    \centering
     \includegraphics[width=\columnwidth]{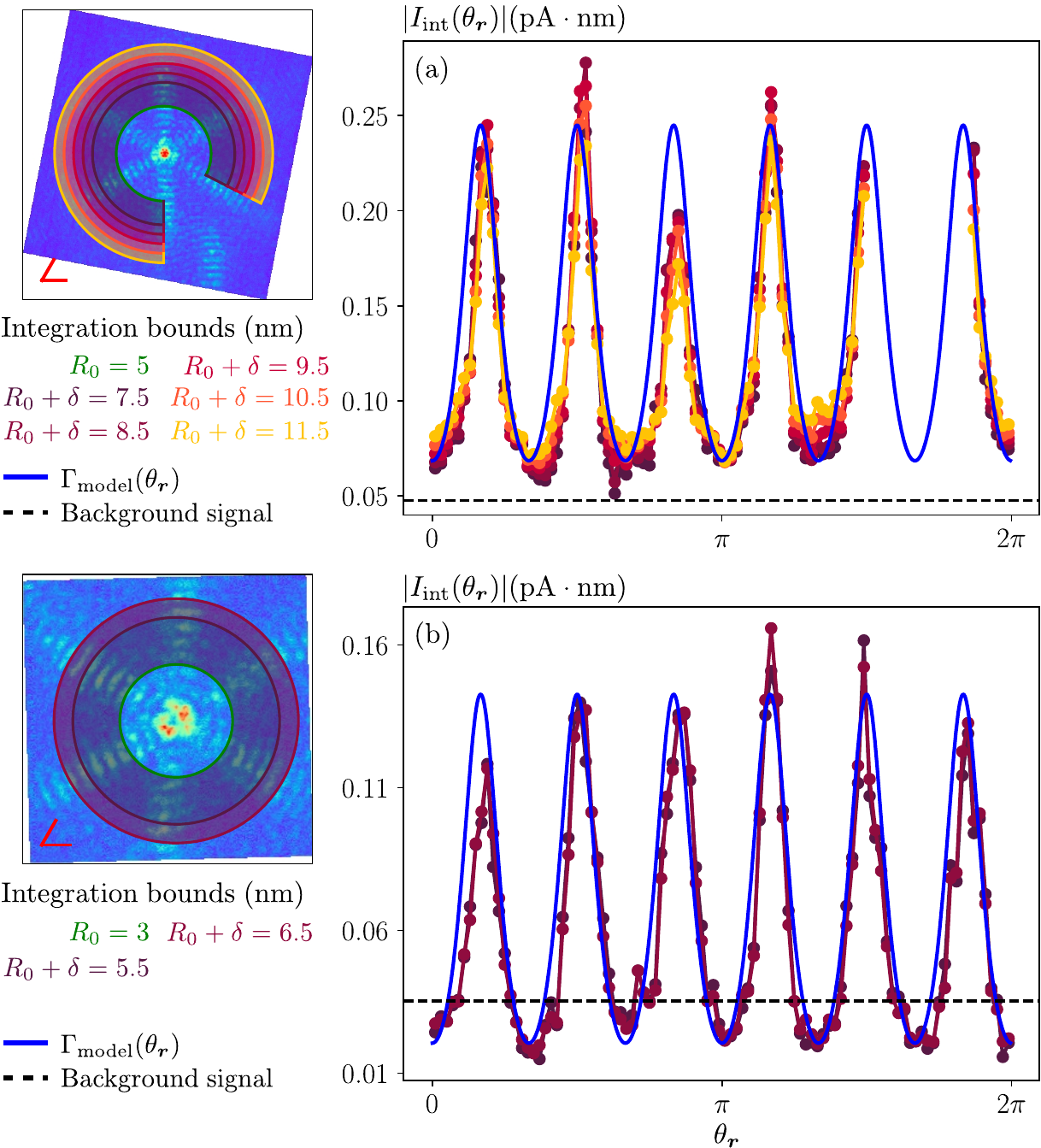}
    \caption{Angular dependence of the LDOS prefactor. Panels (a) and (b) correspond to data sets 1 and 2, respectively. Purple-to-yellow curves represent experimental data points $|I_{\mathrm{int}}|(\theta)$ for $\delta$ values specified in the legend and $d\theta= 2\pi/100$. Legend inset shows the real-space current data with a lattice vector aligned with the horizontal axis ($\theta_{\bm{r}}=0$). We excluded a range of data points in data set 1 due to interference with another YSR state (partially visible on the bottom right of the image).}
    \label{fig:pref_analysis}
\end{figure}

In Fig.~\ref{fig:pref_analysis} we present the result of our analysis in comparison with the model curve
\begin{equation}
\label{eq:gamma_model}
\Gamma_{\mathrm{model}}(\theta_{\bm{r}}) = a \Gamma_{\mathrm{max}}(\theta_{\bm{r}}) + a_0.
\end{equation}
Here $a$ is a free parameter in the theory, and $a_0$ is constant shift on the order of the background signal. We estimate the latter by taking the signal average far away from the impurity times the typical integration length $\overline{\bm{R}}_n \sim R_0 + \frac{\delta_{\mathrm{min}}+\delta_{\mathrm{max}}}{2}$.

Experimental data points $\rho_{\mathrm{int}}^{R_0,\delta}(\theta)$ above the background signal collapse onto a universal curve, thereby supporting the assumptions stated in the previous paragraph.

The relative angular position of the model curve with respect to the data points is not a degree of freedom, but it is determined by the orientation of the lattice vectors (see legend insets in Fig.~\ref{fig:pref_analysis}), the latter being extracted from the constant-current image of the atomic surface. We emphasize the agreement between the positions of the maxima and minima of the model and the experimental curves. While the six-fold nature of the YSR state may follow trivially from the symmetry of the band structure, its orientation with respect to the lattice does not. One could generally expect YSR states to be oriented according to high-symmetry lines; however, it is a priori equally reasonable for the ``petals'' of the LDOS to be along the lattice vectors or in between. The theory shows how this is, in fact, determined by the geometry of the Fermi surface and Fermi velocity.

We note that the periodicity of the data points with $\theta_{\bm{r}}$ exhibits certain higher harmonics, namely, the signal becomes thinner at the crests and wider at the valleys, in other words, the ``petals'' are narrow. This nontrivial feature is also present in the model curve.

\subsection{Fourier-space analysis}
\label{ssec:fourier_space}

\begin{figure}
    \centering
     \includegraphics[width=\columnwidth]{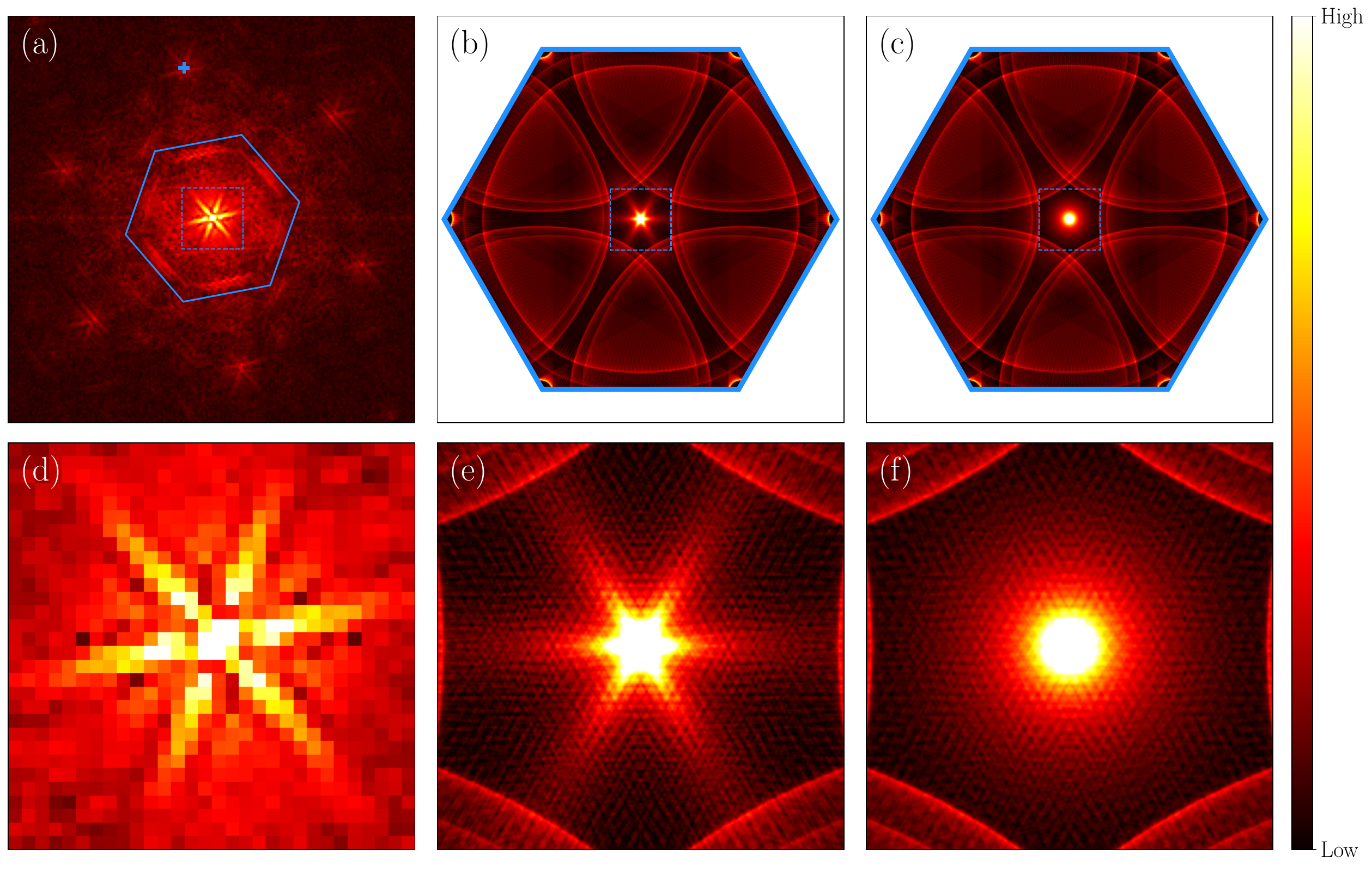}
    \caption{(a) Fourier transform's modulus of the in-gap current shown in Fig.~\ref{fig:spatial_data} (b). The blue hexagon indicates the FBZ and the blue cross marks one of the Bragg peaks. (b) Fourier transform's modulus of the analytical expression of the LDOS [Eq.~\eqref{eq:rho_fft}] evaluated on the FBZ. (c) Same as (b) with constant prefactor. (d), (e), (f) Zoom around the origin delimited by the dashed-line square in panels (a), (b),  and (c), respectively.}
    \label{fig:raw_fft}
\end{figure}

To further confirm the validity of our interpretation, we study the Fourier transform (FT) of the in-gap recorded current [Fig.~\ref{fig:raw_fft} (a)], and we compare it against the FT of the asymptotic expression of the LDOS, which contains the prefactor, the exponential decay, and the oscillations,
\begin{equation}
    \label{eq:rho_fft}
    \widetilde{\rho}(\bm{q}) = \sum_{j,j'}\int_{0}^{2\pi}d\theta_{\bm{r}} \Gamma_{j,j'}(\theta_{\bm{r}})\widetilde{F}_{j,j'}(\bm{q}, \theta_{\bm{r}}),
\end{equation}
with $\widetilde{F}_{j,j'}(\bm{q}, \theta_{\bm{r}}) \equiv \int dr e^{-r/\xi_{j,j'}(\theta_{\bm{r}})}f[(\bm{k}^{\pm}_{j,j'}(\theta_{\bm{r}})-\bm{q})\cdot\bm{r}]$ the integral over the radial coordinate $r$ (see Appendix~\ref{appFT} for details).

The experimental FT of the YSR states exhibits two salient features that are qualitatively captured by the model [cf.~Figs.~\ref{fig:raw_fft}(a), \ref{fig:raw_fft}(d) and \ref{fig:raw_fft}(b), \ref{fig:raw_fft}(e)]. The first comprises the linear segments parallel to the edge of the hexagonal first Brillouin zone (FBZ), corresponding to the short wavelength oscillations of the YSR state. The second is the six-fold star at small $\bm{q}$ whose orientation is rotated by $\pi/6$ with respect to the YSR state in real-space (cf.~Fig.~\ref{fig:spatial_data}) as expected from the fact that a focused signal in real-space gives a focused feature in Fourier space that is in orthogonal direction. The small star in momentum space is the manifestation of the QFE at large length scales. Finally, we corroborate that the long-range anisotropy of the YSR state, i.e., the small-$\bm{q}$ star feature in Fourier space, is controlled by the prefactor in the asymptotic expression of the LDOS: we set $\Gamma_{j,j'}(\theta_{\bm{r}}) = 1$ in Eq.~\eqref{eq:rho_fft} and obtain a FT map which differs from the full $\widetilde{\rho}(\bm{q})$ exactly in that it lacks the small-$\bm{q}$ sixfold star [Figs.~\ref{fig:raw_fft}(c), \ref{fig:raw_fft}(f)].

\section{Conclusions}
\label{sec:conclusions}
We show that the theory of the quasiparticle focusing effect in two-dimensional superconductors \cite{uldemolinsQuasiparticleFocusingBound2022} can be further simplified in application to STM data of superconductors such as 2H-NbSe$_2$, in which the coherence length derived from the band structure exceeds the spatial extent of YSR states.
Our conclusions are established both by a practical analysis of annuli in real space, and by analysis of features in Fourier space.

The key physical quantity for all practical purposes becomes the prefactor $\Gamma(\theta_{\bm{r}})$ of the LDOS expression, which faithfully captures the symmetry, orientation, and angular profile of YSR states observed in STM experiments. The prefactor has a simple analytical expression depending on the Fermi velocity and curvature of the Fermi surface only. In particular, the LDOS around a magnetic impurity is preferentially focused in directions perpendicular to flatter sections of the Fermi contour, and in the small-gap limit it matches the QFE in the underlying normal metal \cite{lounisTheoryRealSpace2011,weismannSeeingFermiSurface2009}. It would be interesting to test this matching in future measurements of local response to impurities in the normal metallic state of NbSe$_2$.

In conclusion, this work demonstrates the applicability of asymptotic approximations in the spirit of \cite{lounisTheoryRealSpace2011,uldemolinsQuasiparticleFocusingBound2022,weismannSeeingFermiSurface2009} for describing the local response to point defects, and underlines the importance of the band structure geometry for designing collective impurity states.

\begin{acknowledgments}
M.U., A.M., and P.S.~acknowledge the support of the French Agence Nationale de la Recherche (ANR), under Grant No.~ANR-22-CE30-0037.
M.U.~acknowledges funding by the European Union NextGenerationEU/PRTR-C17.I1 and the Department of Education of the Basque Government (IKUR Strategy). F.M.~acknowledges L.~Cario for providing the samples and funding from H2020 Marie Sk\l{}odowska-Curie Actions (Grant No.~659247) and the ANR (Grant No.~ANR-16-ACHN-0018-01).
\end{acknowledgments}
\newpage

\appendix
\begin{widetext}

\section{Experimental details and additional data}
\label{appExp}
2H-NbSe$_{2}$ single crystals were grown using an iodine transport method and were unintentionally doped by a few tens of ppm of magnetic species (Fe, Cr, Mn) contained in the niobium precursor; see Ref. \cite{menardCoherentLongrangeMagnetic2015} for more details. The crystals were mechanically cleaved in cryogenic vacuum at T $\sim$ 20~K and directly inserted into the STM head \cite{massee_revsciinstrum_2018} at 4.2~K. An etched atomically sharp and stable tungsten tip was used for all measurements, which were performed at $T=0.3$~K. In Fig.~\ref{fig-s:spec} we present the $dI/dV$ spectra for the two impurities considered in this work and in Figs.~\ref{fig-s:prefactor_analysis}(b), \ref{fig-s:prefactor_analysis}(d) we analyze the anisotropy of the signal recorded at positive bias to complement the data presented in the main text. We note that the spatial maps presented in this work show $I$ (as opposed to $dI/dV$), but recorded in the fully gapped region between 0 mV and the coherence peak, such that only the signal from the sub-gap state is present.

\begin{figure}[H]
    \centering
     \includegraphics[width=0.6\columnwidth]{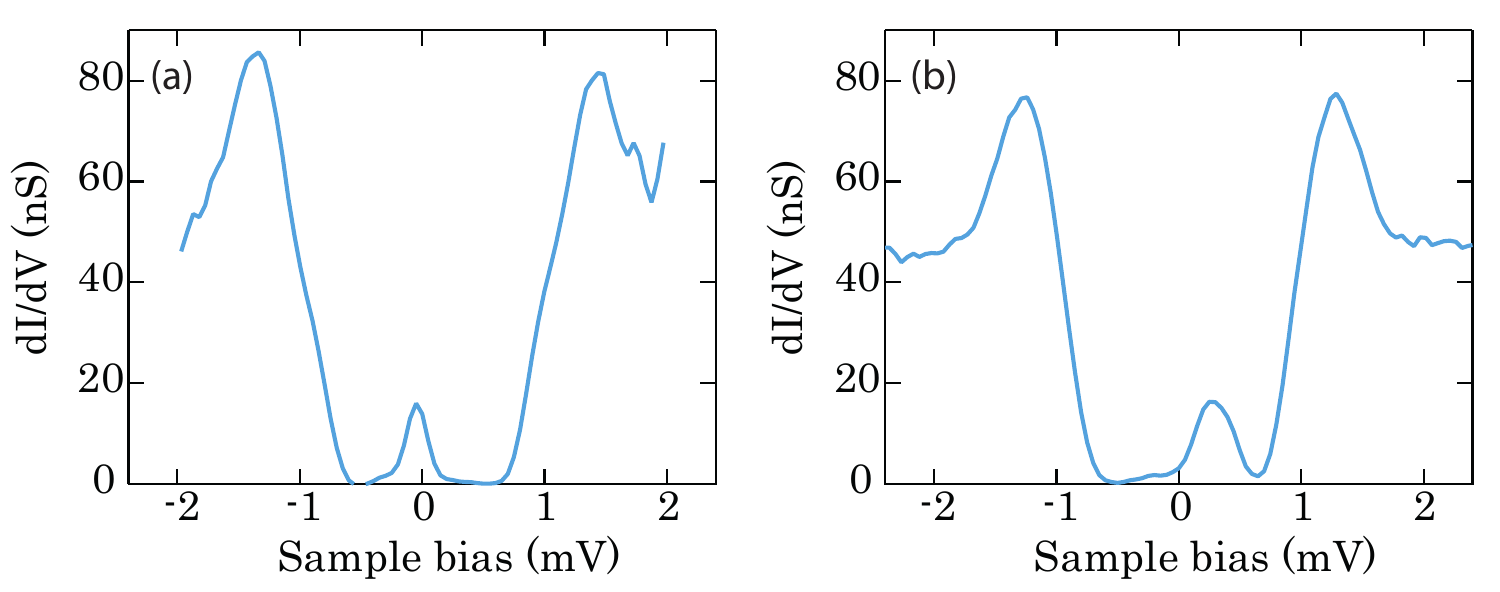}
    \caption{Tunneling spectrum taken at the core of the YSR state in Figs. 1(a), 1(b) (a) and Figs. 1(c), 1(d) (b) showing the strongest in-gap peak at $V_{\mathrm{S}}=-0.1$ meV and $V_{\mathrm{S}}=0.25$ meV, respectively.}
    \label{fig-s:spec}
\end{figure}

\begin{figure}[H]
    \centering
     \includegraphics[width=\columnwidth]{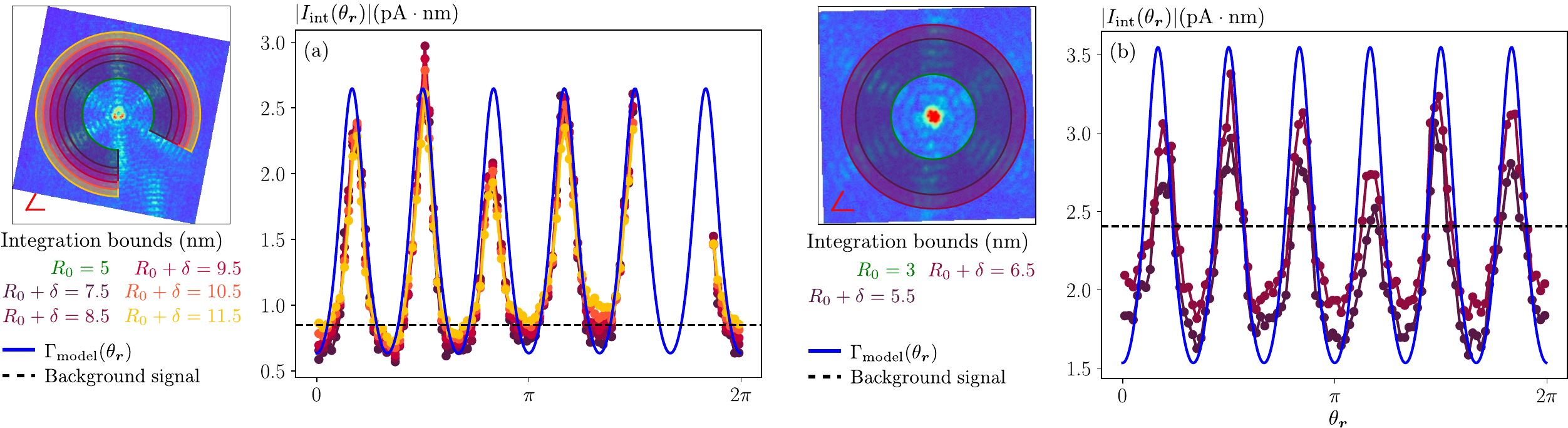}
    \caption{Same analysis of the LDOS prefactor as in Fig.~\ref{fig:pref_analysis} of the main text for the in-gap current recorded at positive bias, $V=0.6$ mV. Panels (a) and (b) correspond to data sets 1 and 2, respectively. The phenomenological parameters in the model curve [Eq.~\eqref{eq:gamma_model}] for panels [Figs.~\ref{fig:pref_analysis}(a), \ref{fig:pref_analysis}(b), and Figs.~\ref{fig-s:prefactor_analysis}(a), \ref{fig-s:prefactor_analysis}(b)] are: $a_0 = (-3,-3,-2.5,-2.1)\cdot 10^{-12}$ pA $\cdot$ nm and $a/a_{\mathrm{c}}=(4.45, 4.45, 3.2, 4.45) \cdot 10^{-7}$ pA $\cdot$ nm $\cdot$ meV$^{1/2}$, with the model curve computed using Eqs.~\eqref{eq-s:ek} to \eqref{eq-s:curv} and ``Band 2'' parameters in Table \ref{tab:ek_param}.}
    \label{fig-s:prefactor_analysis} 
\end{figure}

\section{Derivation of the LDOS expression}
\label{appLDOS}
We derive the expression for the in-gap LDOS in Eq.~\eqref{eq:rho}, which is equal in value to the main result from \cite{uldemolinsQuasiparticleFocusingBound2022}, but is further transformed to make explicit that LDOS is real, and that the doubling of critical momenta in the superconductor with respect to the normal metal does not essentially alter the prefactor. We further show that the prefactor up to linear order in $\Delta/\varepsilon_\textrm{F}$ matches the one of the normal metal.

The bare propagator at energy $E<\Delta$ from $\bm{r}_a=\bm{r}$ to $\bm{r}_b=\bm{0}$ on an s-wave superconductor (described by the Hamiltonian in Eq.~\eqref{eq:h0} of the main text) in Nambu space can be approximated in the far-field limit ($rk_{F, \mathrm{min}} \gg 1$, with $k_{F, \mathrm{min}}$ the minimum Fermi wave vector of the Fermi contour) as 

\begin{equation}
    \label{eq-s:g0_complex}
    \hat{G}_0(\bm{r},\bm{0};E) \sim \frac{1}{\sqrt{r}}\sum_{j,\epsilon=\pm}\Gamma_{j,\epsilon}(\theta_{\bm{r}})\exp\left\{-\frac{r}{\xi_{j,\epsilon}(\theta_{\bm{r}})} +i\left[\bm{r}\cdot \operatorname{Re}\left[\bm{k}_{j,\epsilon}(\theta_{\bm{r}})\right]+\epsilon\frac{\pi}{4}\right]\right\}\begin{pmatrix}
        E+i \epsilon \omega & \Delta\\
        \Delta & E-i\epsilon\omega
        \end{pmatrix},
\end{equation}

where $\theta_{\bm{r}}$ is the polar angle on the plane defined by the vector $\bm{r}$ from the impurity to the point where we seek to evaluate the LDOS, $\omega = \sqrt{\Delta^2-E^2}$ and
\begin{subequations}
    \begin{align}
\Gamma_{j,\epsilon}(\theta_{\bm{r}}) &= \frac{1}{|\nabla\varepsilon_{\bm{k}_{j,\epsilon}(\theta_{\bm{r}})}|\sqrt{\kappa_{\bm{k}_{j,\epsilon}(\theta_{\bm{r}})}}}, \label{eq-s:pref_complex}\\
\xi_{j,\epsilon}(\theta_{\bm{r}}) &= \frac{1}{\operatorname{Im}\left[\bm{k}_{j,\epsilon}(\theta_{\bm{r}})\right]\cdot \hat{\bm{r}}}, \label{eq-s:xi_complex}
    \end{align}
\end{subequations}
with $|\nabla\varepsilon_{\bm{k}_{j,\epsilon}(\theta_{\bm{r}})}|$ norm of the gradient of the energy dispersion $\varepsilon_{\bm{k}}$, and $\kappa_{\bm{k}_{j,\epsilon}(\theta_{\bm{r}})}$ the curvature of the Fermi contour $\varepsilon_{\bm{k}}=0$ evaluated at $\bm{k}_{j,\epsilon}(\theta_{\bm{r}})$. The points $\bm{k}_{j,\epsilon}(\theta_{\bm{r}})$ are the solutions to the saddle-point equations, namely complex momenta satisfying
\begin{subequations}
    \begin{align}
    \varepsilon_{\bm{k}_{j,\epsilon}(\theta_{\bm{r}})} &= i \epsilon \omega, \label{eq-s:saddle_eqs_complex-1}\\
  \frac{\nabla\varepsilon_{\bm{k}_{j,\epsilon}(\theta_{\bm{r}})}}{|\nabla\varepsilon_{\bm{k}_{j,\epsilon}(\theta_{\bm{r}})}|} &= \epsilon \bm{\hat{r}},
    \end{align}
 \end{subequations}
where $j=1,\dots,N$ denote the solutions from multiple disconnected pockets of the Fermi contour, $\epsilon =\pm$, and $\bm{\hat{r}}$ is the unit vector defined by $\theta_{\bm{r}}$. For our present purpose of analyzing a small-gap superconductor, we consider a perturbative expansion of the saddle-point equations in $\Delta/\varepsilon_{\mathrm{F}}$, noting that the normal metal is recovered as the pairing vanishes. This forces two simultaneous limits: (1) $\Delta/\varepsilon_{\mathrm{F}}\rightarrow0$, and (2) $E_S\rightarrow\Delta$, which due to the only in-gap state being at $E=E_S$ effectively means that in Eq.~\eqref{eq-s:saddle_eqs_complex-1} one has $\omega\rightarrow0$.
 In Ref.~\cite{uldemolinsQuasiparticleFocusingBound2022} it was shown that to linear order in $\Delta/\varepsilon_{\mathrm{F}}$, the angle-dependent quantities in the exponent of the propagator \eqref{eq-s:g0_complex} take the following form:
 \begin{subequations}
 \begin{align}
 \operatorname{Re}\left[\bm{k}_{j,\epsilon}(\theta_{\bm{r}})\right] &= \widetilde{\bm{k}}_{j,\epsilon}(\theta_{\bm{r}}),\\
 \xi_{j,\epsilon}(\theta_{\bm{r}}) &= \frac{|\nabla\varepsilon_{\widetilde{\bm{k}}_{j,\epsilon}(\theta_{\bm{r}})}|}{\omega}, \label{eq-s:xi_real}
 \end{align}
 \end{subequations}
 with $\widetilde{\bm{k}}_{j,\epsilon}(\theta_{\bm{r}})$ the solutions to the saddle-point equations in the $\omega \rightarrow 0 $ limit,
 \begin{subequations}
 \label{eq-s:saddle_eqs_small_gap}
    \begin{align}
    \varepsilon_{\widetilde{\bm{k}}_{j,\epsilon}(\theta_{\bm{r}})} &= 0,\\
  \frac{\nabla\varepsilon_{\widetilde{\bm{k}}_{j,\epsilon}(\theta_{\bm{r}})}}{|\nabla\varepsilon_{\widetilde{\bm{k}}_{j,\epsilon}(\theta_{\bm{r}})}|} &= \epsilon \bm{\hat{r}},
    \end{align}
 \end{subequations}
 that is, real momenta positioned on the Fermi contour. To zeroth order in $\Delta/\varepsilon_{\mathrm{F}}$, it is obvious that the prefactor is real and takes the same form as in the normal-metal propagator, i.e., Eq.~\eqref{eq-s:pref_complex} evaluated at $\widetilde{\bm{k}}_{j,\epsilon}(\theta_{\bm{r}})$,
 \begin{equation}
 \label{eq-s:pref_real}
     \Gamma_{j,\epsilon}(\theta_{\bm{r}}) = \frac{1}{|\nabla\varepsilon_{\widetilde{\bm{k}}_{j,\epsilon}(\theta_{\bm{r}})}|\sqrt{\kappa_{\widetilde{\bm{k}}_{j,\epsilon}(\theta_{\bm{r}})}}}.
 \end{equation}
 The question arises whether there is any non-trivial correction to next-leading order, i.e., linear in $\Delta/\varepsilon_{\mathrm{F}}$. To show that there is no correction, we consider an expansion of the prefactor in terms of a small complex quantity $z$ with $\bm{k}_{j,\epsilon}(\theta_{\bm{r}}) = \widetilde{\bm{k}}_{j,\epsilon}(\theta_{\bm{r}})+z$, knowing that $z\rightarrow0$ as $\Delta/\varepsilon_{\mathrm{F}}\rightarrow0$. The key point now is that the angular profile of the measured YSR state depends on the amplitude of the prefactor, so that we write $\Gamma_{j,\epsilon}(\theta_{\bm{r}})\equiv|\Gamma|\exp(i\Phi)$ and consider the expansion of the real function evaluated for a complex value of its variable, $|\Gamma(z)|\approx\sqrt{\left[\Gamma(k)+\frac{1}{2}\Gamma''(k)\textrm{Re}(z^2)\right]^2+\left[\Gamma'(k)\textrm{Im}(z)+\frac{1}{2}\Gamma''(k)\textrm{Im}(z^2)\right]^2}\approx\Gamma(k)+O(\Delta^2)$, where $k\equiv\widetilde{\bm{k}}_{j,\epsilon}(\theta_{\bm{r}})$ is the real solution in the metal [Eq.~\eqref{eq-s:saddle_eqs_small_gap}]. We used that $\Gamma(k)$ is real, and further assumed that the correction $\textrm{Im}(z)$ is at least linear in $\Delta$, while the $\textrm{Re}(z)$ is at least quadratic, which is reasonable due to the structure of Eq.~\eqref{eq-s:saddle_eqs_complex-1}. We note that these properties of $z$ can be confirmed in the particular example of an elliptic Fermi contour worked out in Ref.~\cite{uldemolinsQuasiparticleFocusingBound2022}. For the second step in the expansion we also assumed $\Gamma(k)\neq0$, which is a condition for the entire saddle-point approach to be valid anyway. Hence the correction of the amplitude of the prefactor is at most second order in $\Delta/\varepsilon_{\mathrm{F}}$, so that up to and including first order the prefactor is calculated as for the normal metal [Eq.~\eqref{eq-s:pref_real}]. We note that the prefactor's phase $\Phi$ can in general be linear in $\Delta/\varepsilon_{\mathrm{F}}$, thereby contributing an additional $\theta_{\bm{r}}$-dependent phase to the bare propagator (and hence the LDOS) whose analysis is beyond the scope of this work.

Continuing now the derivation of the LDOS, to all orders in perturbation theory in the strength of the impurity potential $\hat{V}$, the modification of the LDOS due to a point-like scatterer at $\bm{r}=\bm{0}$ is given by 
\begin{equation}
    \rho(\bm{r},E) = -\frac{1}{\pi}\operatorname{Im}\operatorname{Tr}\left[\hat{M}\hat{G}_0(\bm{r},\bm{0};E)\hat{T}(E)\hat{G}_0(\bm{0},\bm{r};E)\right],
\end{equation}
where $\hat{T}(E) = \hat{V}\left[1-\hat{G_0}(\bm{0},\bm{0};E)\hat{V}\right]^{-1}$ is the T-matrix and $\hat{M} = (\tau_0+\tau_z)/2$ projects out the electron-electron component. Here and in the following $\tau_{0,x,y,z}$ denote the Pauli matrices in Nambu space. Note that the counter-propagator satisfies $\hat{G}_0(\bm{0},\bm{r};E) = \hat{G}_0^{*}(\bm{r},\bm{0};E)$. Since the anisotropy of the LDOS is encoded in the bare propagator \eqref{eq-s:g0_complex}, we simplify by considering a magnetic scatterer only, without non-magnetic scattering amplitude, i.e. $\hat{V} = -J \tau_0$. Further, we neglect the anisotropy of the band structure when evaluating $\hat{G_0}(\bm{0},\bm{0};E)$ in the T-matrix,
\begin{equation}
    \hat{G_0}(\bm{0},\bm{0};E) \equiv \frac{1}{\mathcal{V}}\sum_{\bm{k}}\hat{G}_0(\bm{k},E) \sim -\frac{\pi \nu_0}{\sqrt{\Delta^2-E^2}}\begin{pmatrix}
        E&\Delta\\
        \Delta&E
    \end{pmatrix},
\end{equation}
where $\nu_0$ is the density of states at the Fermi level. Under these assumptions and in the small-gap limit we have
\begin{equation}
    \rho(\bm{r})\sim -\frac{1}{\pi} \operatorname{Im}\frac{1}{r}\sum_{j,j',\epsilon,\epsilon'}\Gamma_{j,j'}^{\epsilon,\epsilon'}(\theta_{\bm{r}})\exp\left\{-\frac{r}{\xi_{j,j'}^{\epsilon,\epsilon'}(\theta_{\bm{r}})} +i\left[\bm{r}\cdot\left(\widetilde{\bm{k}}_{j,\epsilon}(\theta_{\bm{r}})-\widetilde{\bm{k}}_{j',\epsilon'}(\theta_{\bm{r}})\right)+(\epsilon-\epsilon')\frac{\pi}{4}\right]\right\}\frac{N_{\epsilon, \epsilon'}}{D},
\end{equation}
where
\begin{subequations}
\begin{align}
    \Gamma_{j,j'}^{\epsilon,\epsilon'}(\theta_{\bm{r}}) &= \Gamma_{j,\epsilon}(\theta_{\bm{r}})\Gamma_{j',\epsilon'}(\theta_{\bm{r}}),\\
    \xi_{j,j'}^{\epsilon,\epsilon'}(\theta_{\bm{r}}) &= \left(\frac{1}{\xi_{j,\varepsilon}(\theta_{\bm{r}})}+\frac{1}{\xi_{j',\varepsilon'}(\theta_{\bm{r}})}\right)^{-1},\\
    N_{\epsilon, \epsilon'} &= t_0 (\Delta^2+E^2)+2E\Delta t_x + t_0 \omega^2 \epsilon\epsilon'+\\\notag
    &+i \omega (Et_0+\Delta t_x)(\epsilon-\epsilon'),\\
    D&=t_0^2-t_x^2,
\end{align}
\end{subequations}
with $t_0 = 1-\alpha \frac{E}{\omega}$, $t_x=\alpha \frac{\Delta}{\omega}$ and $\alpha = \nu_0 \pi J$, and $\Gamma_{j,\epsilon}(\theta_{\bm{r}})$ and $\xi_{j,\epsilon}(\theta_{\bm{r}})$ as defined in Eqs.~\eqref{eq-s:pref_real} and \eqref{eq-s:xi_real}, respectively. The energy of the in-gap bound states are given by the poles of the T-matrix, i.e. $D=0$, and takes the well known form,
\begin{equation}
    E_{\mathrm{S}}^{\pm}= \pm \Delta \frac{1-\alpha^2}{1+\alpha^2}.
\end{equation}

To obtain the LDOS of the YSR state \cite{rubyTunnelingProcessesLocalized2015}, we evaluate $N_{\epsilon,\epsilon'}$ at $E=E_{\mathrm{S}}^{+}$ and expand $D$ to first order around $E=E_{\mathrm{S}}^{+}$ to obtain up to a constant prefactor,
\begin{equation}
    \label{eq-s:rhoX}
    \rho(\bm{r})\sim -\frac{1}{\pi} \operatorname{Im}\frac{1}{r}\sum_{j,j',\epsilon,\epsilon'}\Gamma_{j,j'}^{\epsilon,\epsilon'}(\theta_{\bm{r}})\exp\left\{-\frac{r}{\xi_{j,j'}^{\epsilon,\epsilon'}(\theta_{\bm{r}})} +i\bm{r}\cdot\left(\widetilde{\bm{k}}_{j,\epsilon}(\theta_{\bm{r}})-\widetilde{\bm{k}}_{j',\epsilon'}(\theta_{\bm{r}})\right)\right\}\frac{X_{\epsilon, \epsilon'}}{E-E_{\mathrm{S}}},
\end{equation}
where $X_{\epsilon, \epsilon'} = (1+\alpha^2)e^{i\varphi_{\epsilon,\epsilon'}}$ with
\begin{equation}
    \label{eq-s:phi}
    \varphi_{\epsilon,\epsilon'} = \begin{cases} 
        0 & \mathrm{if}\; \epsilon=\epsilon', \\
        \epsilon\left[\frac{\pi}{2}+\operatorname{arg}\left(1-\alpha^2+i2\alpha\right)\right] & \mathrm{if}\; \epsilon\neq\epsilon'.
     \end{cases}
\end{equation}
We note that including a nonmagnetic component to the scattering potential yields a different expression for $X_{\epsilon,\epsilon'}$, but crucially with the same dependence on $\epsilon,\epsilon'$.

The expression in Eq.~\eqref{eq-s:rhoX} can be written in a more compact form by noticing that, owing to the evenness of the energy dispersion $\varepsilon_{\bm{k}}$, for each point $\widetilde{\bm{k}}_{j,+}(\theta_{\bm{r}})$, there exists a point $\widetilde{\bm{k}}_{l,-}(\theta_{\bm{r}})$ such that $\widetilde{\bm{k}}_{l,-}(\theta_{\bm{r}}) = - \widetilde{\bm{k}}_{j,+}(\theta_{\bm{r}})$. If the $j$ pocket is centered around the $\Gamma$ point, then $j=l$, but this is not the case in general. The prefactor and characteristic decay length satisfy that $\Gamma_{j,+}(\theta_{\bm{r}})=\Gamma_{l,-}(\theta_{\bm{r}})\equiv \Gamma_{j}(\theta_{\bm{r}})$ and $\xi_{j,+}(\theta_{\bm{r}})=\xi_{l,-}(\theta_{\bm{r}})\equiv \xi_{j}(\theta_{\bm{r}})$, therefore, by relabeling the critical points as $\{\epsilon \widetilde{\bm{k}_j}(\theta_{\bm{r}})\}_{\begin{subarray}{l}j=1\dots N\\ \epsilon=\pm\end{subarray}}$, it becomes obvious that the only contribution to the LDOS comes from the poles of the T-matrix,
\begin{equation}
    \label{eq-s:rho_final}
    \rho(\bm{r}) \sim \frac{1}{r}\sum_{j,j'}\Gamma_{j,j'}(\theta_{\bm{r}}) e^{-r/\xi_{j,j'}(\theta_{\bm{r}})}\sum_{\epsilon,\epsilon'}\exp\left\{i\left[\bm{r}\cdot\left(\epsilon\widetilde{\bm{k}}_{j}(\theta_{\bm{r}})-\epsilon'\widetilde{\bm{k}}_{j'}(\theta_{\bm{r}})\right)+\varphi_{\epsilon,\epsilon'}\right]\right\}\delta(E-E_{\mathrm{S}}),
\end{equation}
where Eq.~\eqref{eq:rho} in the main text is recovered by introducing
\begin{equation}
    f[\bm{k}^{\pm}_{j,j'}(\theta_{\bm{r}})\cdot\bm{r}] \equiv \sum_{\epsilon,\epsilon'}\exp\left\{i\left[\bm{r}\cdot\left(\epsilon\widetilde{\bm{k}}_{j}(\theta_{\bm{r}})-\epsilon'\widetilde{\bm{k}}_{j'}(\theta_{\bm{r}})\right)+\varphi_{\epsilon,\epsilon'}\right]\right\},
\end{equation}
that is a real-valued function. With the new notation the $\epsilon=-$ solutions to Eq.~\eqref{eq-s:saddle_eqs-1} are constructed from the $\epsilon=+$ solutions; therefore, Eq.~\eqref{eq:saddle_eq} in the main text considers the parallel solutions only. Note, however, that the antiparallel solutions are encoded in $ f[\bm{k}^{\pm}_{j,j'}(\theta_{\bm{r}})\cdot\bm{r}]$.

\section{Analysis of the tight-binding energy dispersion}
\label{appTB}
The starting point to compute the model curve for the prefactor (Fig.~\ref{fig:pref_analysis} in the main text) is a tight-binding energy dispersion that describes the two lowest-lying Nb 4d bands of 2H-NbSe$_2$,
\begin{equation}
    \label{eq-s:ek}
    \varepsilon_{\bm{k}} = \mu + \sum_{i = 1}^{5} t_i g_i(\bm{k}),
\end{equation}
where $t_i$ is the hopping amplitude to the $i$-th nearest neighbor and
\begin{subequations}
\begin{align}
  g_1(\bm{k}) &= 2\cos\zeta\cos\eta + \cos 2\zeta,\\
  g_2(\bm{k}) &= 2\cos 3\zeta\cos\eta + \cos 2\eta,\\
  g_3(\bm{k}) &= 2\cos 2\zeta\cos 2\eta + \cos 4\zeta,\\
  g_4(\bm{k}) &= \cos \zeta\cos 3\eta + \cos 5\zeta \cos \eta, + \cos 4 \zeta \cos 2 \eta,\\
  g_5(\bm{k}) &= 2\cos 3\eta\cos3\zeta + \cos 6\zeta,
\end{align}
\end{subequations}
with $\zeta = \frac{1}{2}k_x a_{\mathrm{c}}$ and $\zeta = \frac{\sqrt{3}}{2}k_y a_{\mathrm{c}}$, and $a_{\mathrm{c}}$ is the lattice constant. The hopping amplitudes are presented in Table \ref{tab:ek_param}. We follow Refs.~\cite{menardCoherentLongrangeMagnetic2015,liebhaberYuShibaRusinov2020} and assume that the adatom primarily couples to one of the bands only. Specifically, we take Band 2, which we treat as a spin-degenerate energy dispersion.
 \begin{table}[H]
\centering
\def\arraystretch{1}
\begin{tabular}{rcccccc}
\hline
\multicolumn{1}{l}{} & $\mu$ & $t_1$ & $t_2$ & $t_3$ & $t_4$ & $t_5$ \\ \hline
Band 1               & 10.9    & 86.8  & 139.9 & 29.6  & 3.5   & 3.3   \\
Band 2               & 203.0   & 46.0  & 257.5 & 4.4   & -15.0 & 6.0   \\ \hline
\end{tabular}
\caption{Fitted tight-binding parameters for the two lowest-lying Nb 4d bands of 2H-NbSe\textsubscript{2}. All parameters in meV, for $a_{\mathrm{c}} = 3.444$ \r{A}. Extracted from Ref.~\cite{rahnGapsKinksElectronic2012}}
\label{tab:ek_param}
\end{table}

 \begin{figure}[H]
    \centering
     \includegraphics[width=\columnwidth]{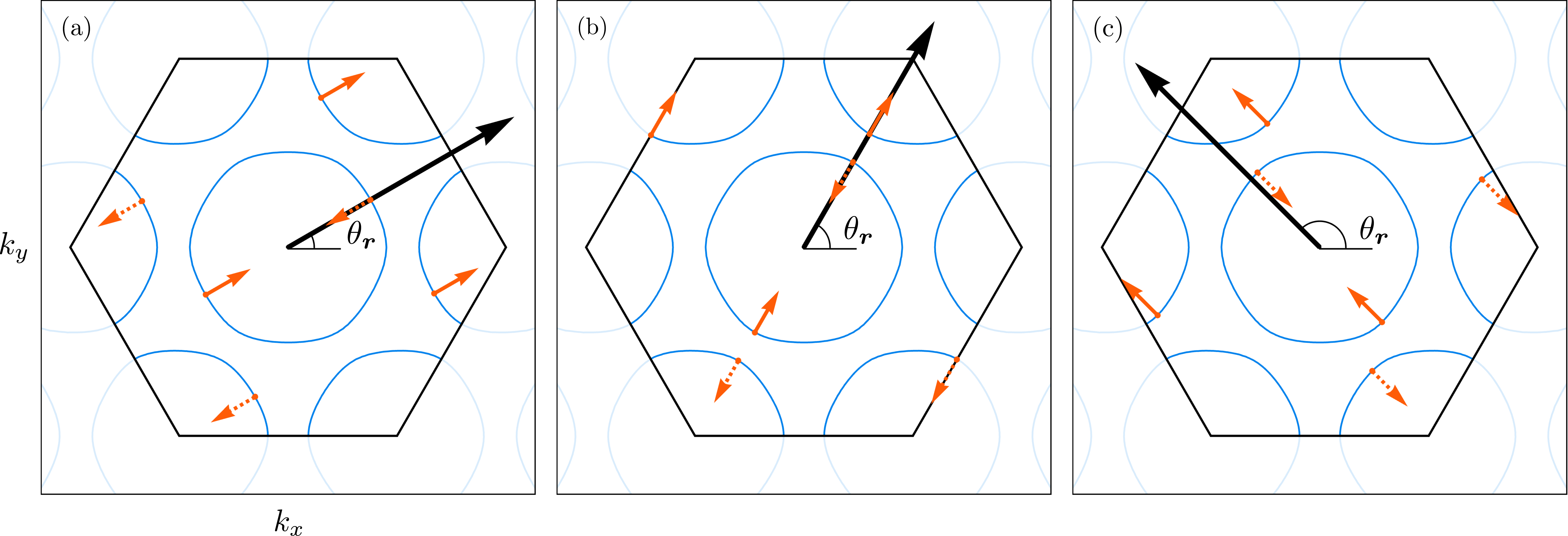}
    \caption{Examples of critical points for various values of $\theta_{\bm{r}}$. The blue lines indicate the Fermi contour for the chosen 2H-NbSe$_2$ model, i.e. Eq.~\eqref{eq-s:ek} with parameters of Band 2. The black arrow represents $\hat{\bm{r}}$, i.e., a vector in real space from the impurity location to an arbitrary position where we evaluate the LDOS. The orange dots on the Fermi contour indicate the corresponding critical points $\epsilon\widetilde{\bm{k}}_j(\theta_{\bm{r}})$, i.e., the solutions to Eqs.~(\ref{eq-s:saddle_eqs-1} and \ref{eq-s:saddle_eqs-2}) ($\epsilon=+$) and their spatially inverted counterparts ($\epsilon=-$). The orange arrows indicate the normalized gradient of the energy dispersion at that point, parallel ($\epsilon=+$) and antiparallel ($\epsilon=-$) to $\hat{\bm{r}}$. Panels (a), (b), and (c) correspond to $\theta_{\bm{r}}=\pi/6, \pi/3, 3\pi/2$, respectively.}
    \label{fig-s:fs}
\end{figure}

The Fermi velocity and the curvature of the Fermi contour that control the anisotropy of the LDOS are related to the energy dispersion as follows,
\begin{subequations}
    \begin{align}
        |\nabla\varepsilon_{\bm{k}}| &= \sqrt{(\partial_{k_x} \varepsilon_{\bm{k}})^2 + (\partial_{k_y}\varepsilon_{\bm{k}})^2},\\
        \kappa_{\bm{k}} &= \frac{(\partial_{k_x} \varepsilon_{\bm{k}})^2 \partial^2_{k_y}\varepsilon_{\bm{k}} + (\partial_{k_y} \varepsilon_{\bm{k}})^2 \partial^2_{k_x}\varepsilon_{\bm{k}} - 2 \partial_{k_x} \varepsilon_{\bm{k}}\partial_{k_y} \varepsilon_{\bm{k}}\partial^2_{k_x,k_y}\varepsilon_{\bm{k}}}{|\nabla\varepsilon_{\bm{k}}|^3}. \label{eq-s:curv}
    \end{align}
\end{subequations}

To compute the model curve for the prefactor we first obtain the critical points in the FBZ by numerically solving
\begin{subequations}
    \begin{align}
    \varepsilon_{\widetilde{\bm{k}}_{j}(\theta_{\bm{r}})} &= 0, \label{eq-s:saddle_eqs-1}\\
  \frac{\nabla\varepsilon_{\widetilde{\bm{k}}_{j}(\theta_{\bm{r}})}}{|\nabla\varepsilon_{\widetilde{\bm{k}}_{j}(\theta_{\bm{r}})}|} &= \bm{\hat{r}}, \label{eq-s:saddle_eqs-2}
    \end{align}
 \end{subequations}
 for a discrete set of $\theta_{\bm{r}}$ values. As the Fermi contour of the energy dispersion in Eq.~\eqref{eq-s:ek} has three non-equivalent pockets, we obtain six critical points $\{\epsilon \widetilde{\bm{k}_j}(\theta_{\bm{r}})\}_{\begin{subarray}{l}j=1,2,3\\ \epsilon=\pm\end{subarray}}$ for a given direction in real space determined by $\theta_{\bm{r}}$ (see Fig.~\ref{fig-s:fs}). This yields a set of six distinct $\Gamma_{j,j'}(\theta_{\bm{r}})$ curves that are essentially in phase [see Fig.~\ref{fig-s:pref} (a)]. Crucially, the prefactor curves are in phase with respect to the characteristic decay length curves as well [Fig.~\ref{fig-s:pref} (b)], which justifies the simplification of taking the dominant prefactor $\Gamma_{\mathrm{max}}(\theta_{\bm{r}}) = \mathrm{max}\{\Gamma_{j,j'}(\theta_{\bm{r}})\}$ in the model curve.

\begin{figure}
    \centering
     \includegraphics[width=\columnwidth]{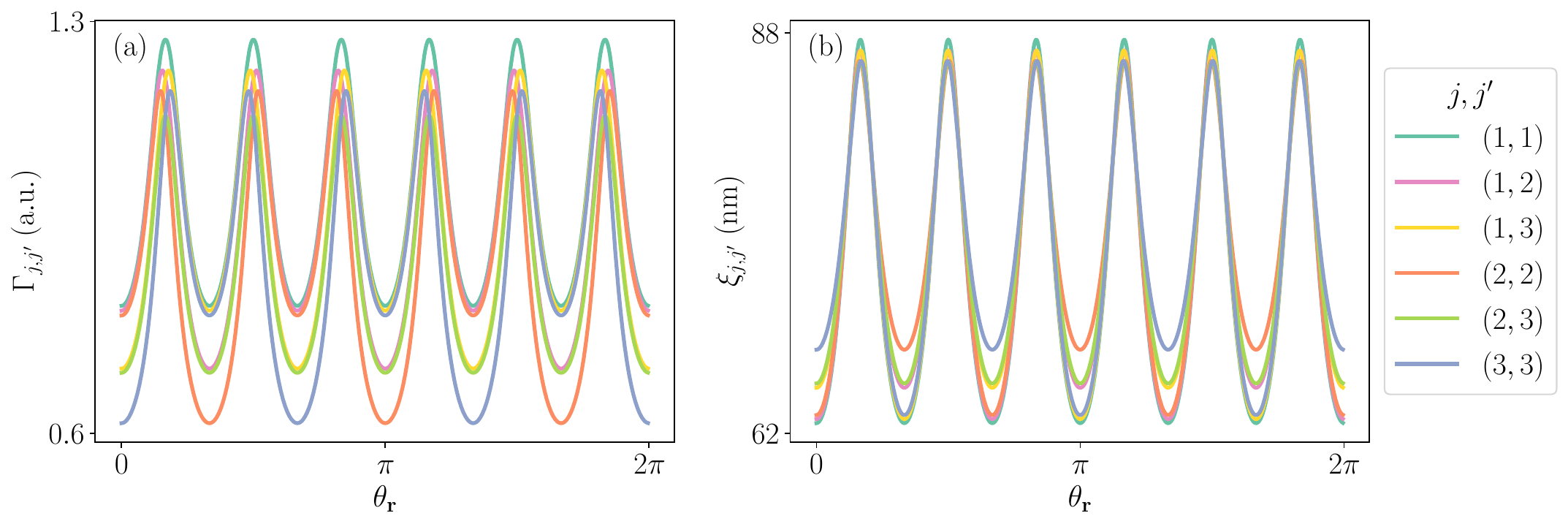}
    \caption{(a) Angular dependence of the LDOS prefactor. (b) Angular dependence of LDOS exponential decay length with $\omega = 1$ meV.  Each curve denotes an interference term $(j,j')$ with $j,j'=1,2,3$ for pockets centered at $\Gamma=(0,0)$, $K=(0,4\pi/3a)$ and $K'=-K$. Certain curves are three-fold periodic only reflecting the reduced symmetry of the $K,K'$ pockets [e.g.~$(2,2)$], yet, the six-fold symmetry of the band structure is recovered when taking into account all the interference terms. The curve presented in Fig.~\ref{fig:pref_analysis} in the main text corresponds to curve $(1,1)$ in panel (a). Compare with Fig.~\ref{fig-s:fs} to observe that both the prefactor and the exponential decay length are maximal when $\hat{\bm{r}}$ is perpendicular to the flattest sections of the Fermi contour.}
    \label{fig-s:pref}
\end{figure}

\section{Effect of Charge Density Wave}
\label{app:CDW}

The presence of charge density wave (CDW) in principle affects two aspects of the YSR LDOS, namely, there is a deformation of the underlying normal state Fermi surface, and there is a modification of the coupling of the impurity to the substrate.

Concerning the effect on the impurity first, Liebhaber and coworkers established in \cite{liebhaberYuShibaRusinov2020} that the relative position of the magnetic adatom with respect to the CDW ordering can significantly modify the properties (e.g.,~reduce the spatial symmetry) of the Yu-Shiba-Rusinov wave function. Impurities studied in this work are subsurface; therefore, it is challenging to determine their exact location with respect to the CDW modulation. However, since the observed LDOS is sixfold symmetric at long-range distances, we can safely assume that the corresponding magnetic impurities are situated on a maximum or a minimum of the CDW, or, alternatively, that the local CDW effect is negligible for buried defects. Hence, we will neglect any influence on the impurity.

Concerning the effect on the band structure, the Fermi surface deformed by the CDW may change the prefactor that is essential to the QFE theory. The conclusion revealed in the rest of this Appendix is that the CDW introduces qualitative changes which may manifest strongly in other compounds, but the magnitude of these effects is negligible in 2H-NbSe$_2$.

We borrow the model for CDW from \cite{liebhaberYuShibaRusinov2020} where they consider an effective mean-field description of the electron-phonon coupling and select the phonon modes corresponding to the lowest Fourier harmonics only. Under these assumptions, the quasi-$(3\times 3)$-commensurate CDW ordering is described by the Hamiltonian
\begin{equation}
    H_{V} = \sum_{\bm{r}}V(\bm{r}) c^{\dagger}_{\bm{r},\sigma}c_{\bm{r},\sigma},
\end{equation}
where $V(\bm{r}) = V_0\sum_{i=1}^3 \cos\left(\bm{Q}_i\cdot \bm{r} + \phi_i\right)$ is a periodic potential with amplitude $V_0 = -30.0$ meV, 
\begin{subequations}
\label{eq-s:cdw_vectors}
\begin{align}
    \bm{Q}_1 &= q(1-\delta)\left(\sqrt{3}/2,-1/2\right),\\
    \bm{Q}_2 &= q(1-\delta)\left(0,1\right),\\
    \bm{Q}_3 &= -\bm{Q}_1-\bm{Q}_2,
\end{align}
\end{subequations}
the CDW wavevectors, and the phases chosen to be $\phi_1=\phi_2 = 0$ and $\phi_3 = -\pi/3$. In Eqs.~\eqref{eq-s:cdw_vectors}, $q = \frac{4\pi}{3\sqrt{3}a_{\mathrm{c}}}$ and $\delta \ll 1$ accounts for the small incommensurability of the CDW with the Nb lattice that we shall neglect. In practice, we apply the method described in App.~\ref{appTB} to each of the six eigenbands at the Fermi energy of the following $9\times9$ Hamiltonian in the basis $\left(c_{\bm{k}}, c_{\bm{k}+\bm{Q}_1}, c_{\bm{k}-\bm{Q}_1}, c_{\bm{k}+\bm{Q}_2}, c_{\bm{k}-\bm{Q}_2}, c_{\bm{k}+\bm{Q}_3}, c_{\bm{k}-\bm{Q}_3}, c_{\bm{k}+\bm{Q}_4}, c_{\bm{k}-\bm{Q}_4} \right)^{T}$,

\begin{equation}
H_{\mathrm{CDW}}=\begin{pmatrix}
\varepsilon(k) & V_1 & V_1^* & V_2 & V_2^* & V_3 & V_3^* & \cdot & \cdot \\
 & \varepsilon(k+\bm{Q}_1) & V_1 & \cdot & V_3 & \cdot & V_2 & V_2^* & V_3^* \\
 &  & \varepsilon(k-\bm{Q}_1) & V_3^* & \cdot & V_2^* & \cdot & V_3 & V_2 \\
 &  &  & \varepsilon(k+\bm{Q}_2) & V_2 & \cdot & V_1 & V_3^* & V_1^* \\
 &  &  &  & \varepsilon(k-\bm{Q}_2) & V_1^* & \cdot & V_1 & V_3 \\
 &  &  &  &  & \varepsilon(k+\bm{Q}_3) & V_3 & V_1^* & V_2^* \\
 &  & \mathrm{c.c.} &  &  &  & \varepsilon(k-\bm{Q}_3) & V_2 & V_1 \\
 &  &  &  &  &  &  & \varepsilon(k+\bm{Q}_4) & \cdot \\
 &  &  &  &  &  &  &  & \varepsilon(k-\bm{Q}_4) 
\end{pmatrix}
\label{eq-s:h_cdw}
\end{equation}
where $\bm{Q}_4 = \bm{Q}_1-\bm{Q}_2$, $\delta=0$ in Eqs.~\eqref{eq-s:cdw_vectors}, $\varepsilon(\bm{k})$ is the energy dispersion defined in Eq.~\eqref{eq-s:ek}, $V_i = \frac{V_0}{18}e^{i\phi_i}$, and $\cdot = 0$, while ``$*$'' and ``$\mathrm{c.c.}$'' denote complex conjugation.

\begin{figure}
    \centering
     \includegraphics[width=0.85\columnwidth]{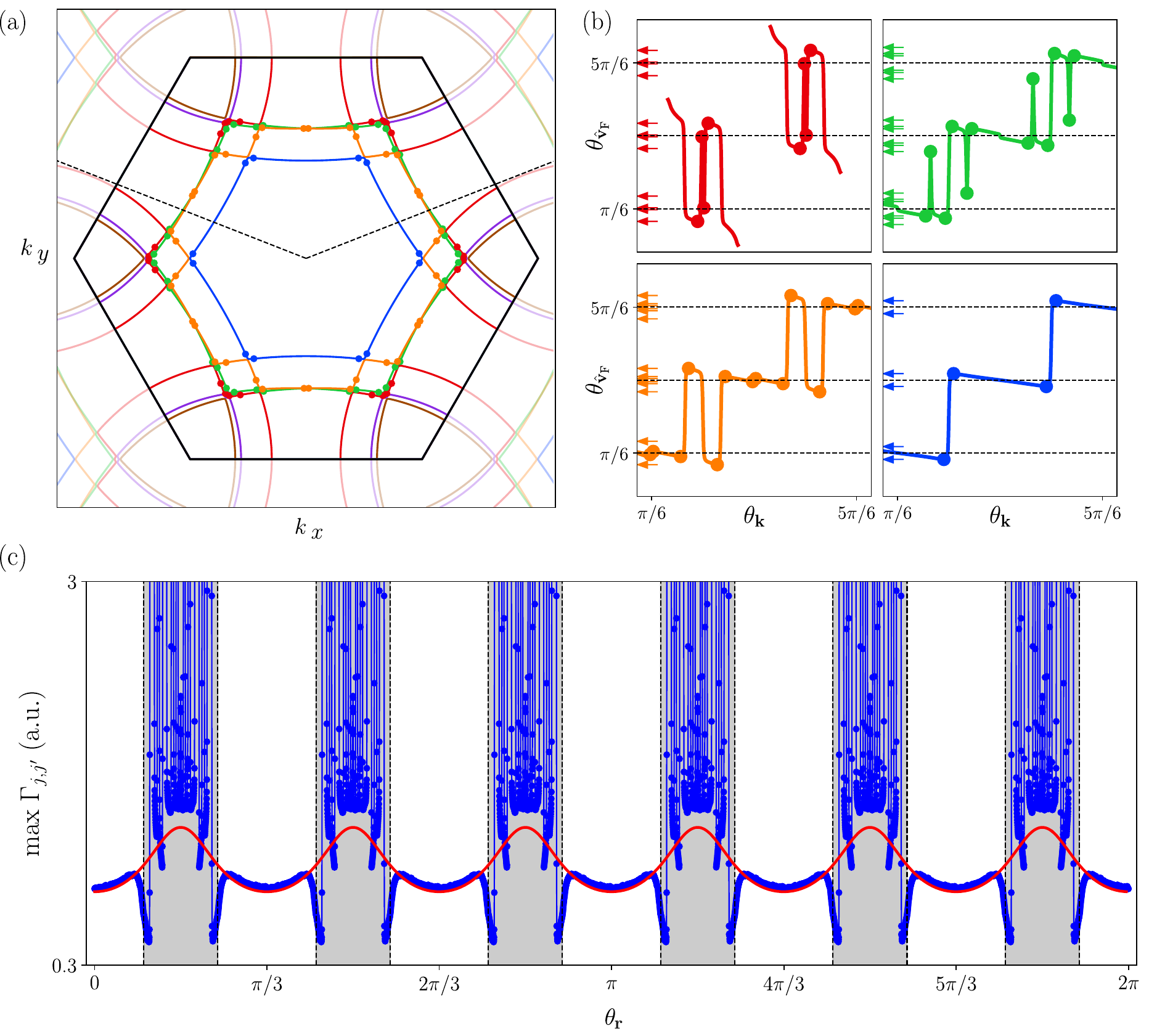}
    \caption{Effect of CDW on the focusing effect in 2H-NbSe$_2$. (a) Fermi contours with a finite CDW potential. Each color represents a band of the Hamiltonian in Eq.~\eqref{eq-s:h_cdw}. The circular markers indicate the inflection points (IPs). The black hexagon delimits the reduced FBZ, which is nine times smaller than the FBZ in Fig.~\ref{fig-s:fs} owing to band folding. (b) Angle of the Fermi velocity $\theta_{\bf{\hat{v}}_{\mathrm{F}}}$ as a function of the polar angle in momentum space $\theta_{\bm{k}}$ for each of the Fermi contours hosting IPs, which are indicated by the circular markers. The small arrows mark the corresponding value $\theta_{\bf{\hat{v}}_{\mathrm{F}}}$ of the IPs (including those outside the shown $\theta_{\bm{k}}$ range). The horizontal dashed black lines indicate the high-symmetry points on which the maxima of the prefactor curve in the absence of CDW potential lie. The horizontal axes span the range delimited by the dashed black radii in panel (a). (c) Angular dependence of the LDOS prefactor. The blue (red) curve corresponds to the calculation with finite (vanishing) CDW potential $V_0$ (the red curve is equivalent to the model curve in Fig.~\ref{fig:pref_analysis}, with $a_0 = 0$ and $a=1$). The black shaded rectangles indicate the $\theta_{\bm{r}}$ regions for which the corresponding critical momenta fall on the vicinity of an IP, and, therefore, the QFE theory is not applicable. In addition to the fundamental break down of the theory, the calculation of Eq.~\eqref{eq:Gamma} in these regions is numerically unstable owing to the vanishing second derivatives, hence the prefactor curve appears to be very noisy.}
    \label{fig-s:cdw}
\end{figure}

If the amplitude of the CDW potential $V_0$ vanishes, the Hamiltonian in Eq.~\eqref{eq-s:h_cdw} is diagonal and follows from a redefinition of the unit cell. The resulting band-folded Fermi contours naturally give rise to the same prefactor as the unfolded band structure. A finite (but small) CDW potential opens small gaps at the intersection of the band-folded Fermi contours \cite{Silva-Guillen_2016}. Due to this hybridization, a pair of inflection points (IPs) with vanishing curvature $\kappa_{\bm{k}}$ appears in the vicinity of the contour-touching points. The pair of IPs emerges in one of the contours involved in the crossing only, namely, the contour which goes from concave to convex and to concave again [see, for instance, the intersection of the $\Gamma$ pocket with itself (blue and orange contours at $\theta_{\bm{k}} = n \pi/3$, $n = 1\dots6$) in Fig.~\ref{fig-s:cdw} (a)]. The QFE theory is not applicable on these singular points since a vanishing curvature requires including next-to-leading order corrections in the saddle point approximation \cite{uldemolinsQuasiparticleFocusingBound2022}, but, in general, one can expect that they create an enhancement of the LDOS prefactor. It is then natural to ask if the emerging inflection points with vanishing curvature yield new peaks in the prefactor curve that would appear as additional petals in the YSR LDOS.

Answering that question boils down to determining the set of real-space polar angles $\theta_{\bm{r}}$ for which the corresponding critical points $\epsilon \widetilde{\bm{k}}_j(\theta_{\bm{r}})$ lay on the IPs of the Fermi contours. If the former differ from the $\theta_{\bm{r}}$ with maximum prefactor at zero CDW potential, new petals will appear in the YSR LDOS. We recall that the Fermi velocity at the critical momenta is parallel to the polar vector along the observation direction in real space $\bm{\hat{r}}$ [see Eq.~\eqref{eq:saddle_eq}], therefore the Fermi velocity angle allows us to relate $\theta_{\bm{r}}$ to $\theta_{\bm{k}}$ for each band [Fig.~\ref{fig-s:cdw} (b)]. The curves $\theta_{\bm{\hat{v}}_{\mathrm{F}}}(\theta_{\bm{k}})$ implicitly contain all the information about the curvature of the Fermi contour since the Fermi velocity vector is perpendicular to the latter by definition. In particular, a positive (negative) slope indicates a convex (concave) segment in the Fermi contour, and, consequently, the extrema of $\theta_{\bm{\hat{v}}_{\mathrm{F}}}(\theta_{\bm{k}})$ correspond to the IPs where the curvature vanishes. The sharp \textit{positive} variations in the $\theta_{\bm{\hat{v}}_{\mathrm{F}}}(\theta_{\bm{k}})$ curve describe the highly curved convex segments of the Fermi contours that result from the CDW hybridization and, consequently, they always appear flanked by two IPs. Further, they coincide with sharp \textit{negative} steps in the $\theta_{\bm{\hat{v}}_{\mathrm{F}}}(\theta_{\bm{k}})$ curve of the other contour involved in the contour-touching point (recall, e.g., the blue and orange contour touching at $\theta_{\bm{k}} = n \pi/3$, $n = 1\dots6$). The sharp negative variations of $\theta_{\bm{\hat{v}}_{\mathrm{F}}}(\theta_{\bm{k}})$ also describe highly curved segments resulting from the CDW hybridization, but in this case, they do not involve any change in the curvature sign, and therefore, there are not any IPs associated with them.

The analysis in Fig.~\ref{fig-s:cdw}(b) demonstrates that, for a choice of $V_0$ relevant to 2H-NbSe$_2$, the $\theta_{\bm{r}}$ corresponding to the the IPs are distributed over a finite region around the high-symmetry points $\theta_{\bm{r}} = (n+1/2) \pi/3$, $n = 1\dots6$ where the theory of the QFE to lowest order of approximation fails. The full calculation of the LDOS prefactor in Fig.~\ref{fig-s:cdw}(c) shows that the prefactor curves with finite and vanishing CDW potential perfectly match within the region of validity. Crucially, the breakdown regions where an enhancement of the LDOS prefactor may be expected are aligned with the maxima of the prefactor curve calculated for $V_0 = 0$. Thus, we can conclude that the band structure modifications induced by the CDW ordering on 2H-NbSe$_2$ do not modify the spatial orientation of the Yu-Shiba-Rusinov states as it was argued above based on the ARPES observations \cite{rahnGapsKinksElectronic2012}.

Finally, it is worth noting that the emergence of inflection points illustrated here is not specific to 2H-NbSe$_2$, but applies to any band structure where CDW can be treated as a commensurate perturbation. In general, a larger value of the CDW potential amplitude $V_0$ will significantly modify the Fermi contours and the Fermi velocity direction at the emergent inflection points may substantially differ from its direction at the lowest-curvature points for $V_0 = 0$, thereby yielding additional focused petals in the YSR LDOS.

\section{Derivation of the Fourier transform of the LDOS}
\label{appFT}
We derive here the Fourier transform of the LDOS at the YSR-state energy:
\begin{equation}
    \begin{split}
    \widetilde{\rho}(\bm{q}) &\equiv \int d\bm{r}e^{-i\bm{q}\bm{r}}\rho(\bm{r})\\
    &= \sum_{j,j'}\int_0^{\infty}dr\int_0^{2\pi}d\theta_{\bm{r}} \Gamma_{j,j'}(\theta_{\bm{r}})\sum_{\epsilon,\epsilon'}\exp\left\{r\left[i\bm{p}_{\epsilon, \epsilon'}(\bm{q},\theta_{\bm{r}})\cdot \hat{\bm{r}}-\frac{1}{\xi_{j,j'}(\theta_{\bm{r}})}\right]\right\}\\
    &= \sum_{j,j'}\int_0^{2\pi}d\theta_{\bm{r}} \Gamma_{j,j'}(\theta_{\bm{r}})\sum_{\epsilon,\epsilon'} e^{i\varphi_{\epsilon,\epsilon'}}\frac{1}{i\bm{p}_{\epsilon, \epsilon'}(\bm{q},\theta_{\bm{r}})\cdot \hat{\bm{r}}-1/\xi_{j,j'}(\theta_{\bm{r}})},
    \end{split}
\end{equation}
where $\bm{p}_{\epsilon, \epsilon'}(\bm{q},\theta_{\bm{r}}) = \epsilon\widetilde{\bm{k}}_{j}(\theta_{\bm{r}})-\epsilon'\widetilde{\bm{k}}_{j'}(\theta_{\bm{r}})-\bm{q}$, and Eq.~\eqref{eq:rho_fft} in the main text is recovered by setting
\begin{equation}
\widetilde{F}_{j,j'}(\bm{q}, \theta_{\bm{r}}) \equiv \sum_{\epsilon,\epsilon'} e^{i\varphi_{\epsilon,\epsilon'}}\frac{1}{i\bm{p}_{\epsilon, \epsilon'}(\bm{q},\theta_{\bm{r}})\cdot \hat{\bm{r}}-1/\xi_{j,j'}(\theta_{\bm{r}})}.
\end{equation}

The maps in Figs.~\ref{fig:raw_fft}(b) and \ref{fig:raw_fft}(c) are obtained by replacing the angular integral by a sum over a discrete set of $\{\theta_{\bm{r}}\}$ values, for each of which the term $\sum_{j,j'} \Gamma_{j,j'}(\theta_{\bm{r}}) \widetilde{F}_{j,j'}(\bm{q}, \theta_{\bm{r}})$ is evaluated. Importantly, to ensure that $\widetilde{\rho}(\bm{q})$ has the periodicity of the reciprocal lattice, the pairs of critical points $\epsilon\widetilde{\bm{k}}_{j}(\theta_{\bm{r}})-\epsilon'\widetilde{\bm{k}}_{j'}(\theta_{\bm{r}})$ have to be mapped onto the FBZ.
Further, as it was shown in Appendix~\ref{appLDOS}, the $\varphi_{\epsilon,\epsilon'}$ phase depends on the nature of the scatterer. For simplicity, we set $\alpha=1$ in Eq.~\eqref{eq-s:phi}, which assumes the strong-coupling regime, i.e., $E_{\mathrm{S}}=0$. 

\end{widetext}

\bibliography{main.bib}

\end{document}